\documentclass[12pt]{article}
\usepackage{amsmath,amssymb,amsfonts,epsf,epsfig,amsthm,bm,graphicx}
\usepackage[dvips]{color}

\begin{document}

\title{
Chern-Simons foam}
\author{Steven Willison and Jorge Zanelli \\
\emph{Centro de Estudios Cient\'{\i}ficos (CECS), Casilla 1469, Valdivia,
Chile}}
\date{3 November 2008}

\maketitle

\begin{abstract}
Chern-Simons theory can be defined on a cell complex, such as a network of
bubbles, which is not a (Hausdorff) manifold. Requiring gauge invariance
determines the action, including interaction terms at the intersections, and
imposes a relation between the coupling constants of the CS terms on
adjacent cell walls. We also find simple conservation laws for charges at
the intersections.
\end{abstract}

{\small CECS-PHY-08/14}

email: steve-at-cecs.cl, z-at-cecs.cl


\section{Introduction}


Soap bubbles have attracted the attention of physicists and mathematicians
for a long time \cite{Boys}. An ordinary bubble is a simple structure
defined by a few elementary rules that can be derived from a minimising
principle: it is a surface of minimal area given a certain constraint --a
fixed enclosed volume or a fixed contour or boundary. These elementary
structures are also the basis for building up more complex structures, like
foams. These aggregates of bubbles also obey some simple rules but include
some nontrivial discrete topological features which must be taken into
account if one tries to derive their laws from an extremal principle.


\subsection{Bubbles and foam}


Perhaps the most celebrated results in the physics of bubbles and foam are
Plateau's rules for minimum area surfaces. A popular example illustrating
these rules are soap films, which, due to surface tension, always tend to
form a shape which minimises the surface area. The rules for a bubble
network are: i) There are a finite number of pieces of film with smooth
curvature, joining at surfaces of smooth intrinsic curvature; ii) The
joining can occur in two ways: either three films meet along a smooth curve,
or four edges (and six films) meet at a point; iii) When three films meet at
a curve, the angle between them is $120^o$. When four edges meet at a point,
the angle at each corner is always a fixed value, given approximately by $%
109^o$. Although it has always been assumed that i)-iii) are
consequences of the minimum area principle, a rigorous proof of this
appeared only fairly recently\cite{Taylor-76}.

The Double bubble theorem is a classic mathematical problem: A
minimal surface containing two adjoining cells of unequal volume is
composed of
three films, each of which is a section of a sphere (Fig. \ref%
{Double_Bubble_Diagram}). The relationship between the three radii of
curvature is

\begin{equation}\label{Double_Bubble_Formula}
\frac{1}{r_1} + \frac{1}{r_2} = \frac{1}{r_3}\, ,
\end{equation}
where films $1$ and $2$ are curved in the opposite direction to film
$3$. In other words the central film which forms the dividing wall
is curved away from the smaller cell into the larger one. The part
of the conjecture that remained unsolved until very recently
\cite{Hutchings-02} is to prove that each of the three films must be
a piece of a sphere. The curvature rule
(\ref{Double_Bubble_Formula}) then follows from this by application
of the $120^o$ rule.

\begin{figure}[t]
\begin{center}
\includegraphics[height=2.65in]{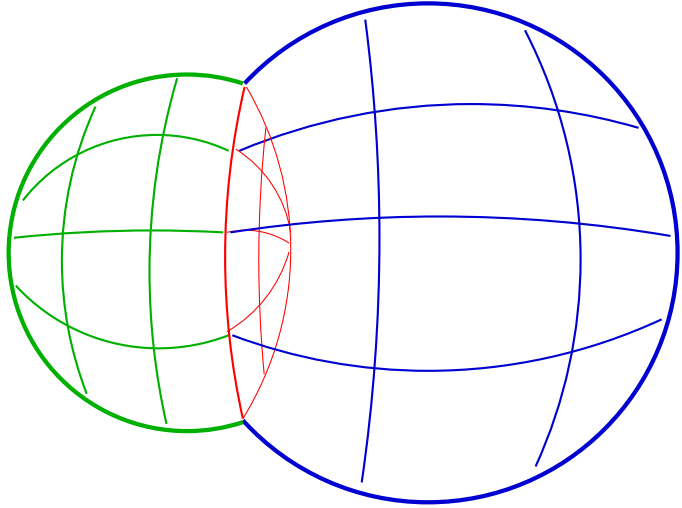}\\[0pt]
\end{center}
\caption{{\protect\small {The double bubble is a familiar structure which
can be seen in soap bubbles. The diagram shows three soap-films joining at a
common edge: the two outer walls (blue and green) and one inner wall (red).
The soap films are two-dimensional. The generalisation to a double bubble
with three-dimensional films could be the base space for a Chern-Simons
theory. Such a structure is not a Hausdorff manifold.}}}
\label{Double_Bubble_Diagram}
\end{figure}



\subsection{Gauge theory of foam?}


Many of the spaces which occur in nature, such as foams and cell structures,
are not manifolds. Or to be precise, the useful approximation (of zero
thickness intersecting films or cell walls) which is often employed, to
simplify their study whilst capturing the essential features, means that one
is not studying a manifold. So, at least at this level of approximation,
non-manifold structures do exist. In this paper, we shall explore the more
speculative possibility that the laws of physics themselves can be
formulated on space-times which are not manifolds, but which are made up of
several manifolds patched together. This is suggested by the special
properties of Chern-Simons (CS) gauge theories, which make them amenable to
a formulation on cell complexes. One can even conceive of a foam made up of
three-dimensional pieces embedded into four dimensions or which has
effective (Hausdorff) dimension four. In this way it may be possible to make
contact between three-dimensional CS theory and four-dimensional physics.
This is an attractive idea as it is widely accepted that at least three of
the four interactions of nature are well described by gauge theories and CS
provides one of the simplest and most elegant gauge theories that we know
of. Unlike most field theories, CS theories demand very little of the
spacetimes on which they can be constructed: the CS action doesn't even need
to have a metric defined on the spacetime manifold.

Here we wish to study Chern-Simons theories constructed on
non-manifold structures. Roughly speaking, we want to investigate
under what conditions a ``Chern-Simons foam" could be consistently
defined. Instead of minimising the surface area, the foam will be
classically described by the extrema of a the topological CS
lagrangian. We will aim to identify simple rules, the analogues of
Plateau's rules or equations such as (\ref{Double_Bubble_Formula}).

The correspondence between CS theory and gravity in three dimensions
suggests that a space-time foam could be described this way. It is perhaps
over-optimistic to think that a foam made from pieces of three-dimensional
gravity will somehow reproduce four-dimensional gravity, but it may be a
hypothesis worth considering. In fact, many current approaches to quantum
gravity involve the breakdown of the manifold structure at some small
length-scales. In any case, even if it turns out that gravity can not be
described in this way, the study is well-motivated from a theoretical point
of view: if CS theory is consistent on non-manifold spaces, it would be
artificial and in some sense unnatural to restrict oneself to a theory
defined only on a manifold.


\subsection{The Chern-Simons three-form}


Before going into details of what kind of non-manifold structures we can
consider, let us review some basic features of the CS three-form for a gauge
field $A$: \cite{Chern-74,Deser-83,Jones-Polynomial,Baez}

\begin{gather}
\mathcal{C}(A) \equiv \text{Tr}\left(A \wedge dA + \frac{1}{3} A
\wedge [A,A]\right)\, .
\end{gather}

Here the gauge field is a one-form which takes values in the Lie algebra $%
\mathcal{G}$ of some gauge group $G$. We use Tr to represent an invariant
bilinear form of the Lie algebra, not necessarily the matrix trace. The Lie
bracket of two exterior differential forms, in this case $[A,A]$, is defined
as $[A,A] \equiv A^a\wedge A^b [J_a , J_b]$.

Some remarks are in order: i) $\mathcal{C}(A)$ is an exterior three-form
that defines the action if integrated over a suitable $3D$ space. If we were
to be conservative, we would insist on a paracompact, oriented topological
manifold. Since the CS form contains first derivatives, we would further
insist that space-time has a differentiable structure, i.e. it must be a
differentiable manifold whose coordinate charts have overlap maps at least
once-differentiable. If we were to define some pathological space, $P$,
which is not a $C^1$ manifold and just naively write down the action $I_P =
\int_P\mathcal{C}$, one might worry that we are doing something ill-defined.
We will need to relax these conservative requirements, but not in any
arbitrary way: fortunately some aspects of the calculus can be generalised
to more general types of spaces, through the notion of integration on \emph{%
chains}. Stokes' theorem is mathematically well-defined on chains and we
shall see that this is sufficient for our purposes.

ii) Physically, the CS gauge theory is unusual in that its ``kinetic term"
is linear in the first derivatives. In the case of a non-abelian gauge
group, there is also a cubic ``self-interaction term". As is well known, in
all cases this theory has no local propagating degrees of freedom except at
the boundary\footnote{%
See e.g. \cite{BGH} for a review including the case of higher dimensional CS
theories, where the phase space is infinite dimensional.}. In fact, the
classical field equation is $F=0$, and therefore all classical solutions are
locally pure gauge. After imposing gauge fixing (e.g. $A_0 =0$) and applying
the constraints, the reduced phase space is the space of flat connections on
the two-dimensional spacelike slice and therefore finite dimensional\cite%
{Jones-Polynomial}.

iii) The CS action $I_{\text{CS}} = (k/4\pi )\int_M \mathcal{C}(A)$ defined
over a manifold without boundary is invariant under infinitesimal gauge
transformations $A \to A+D\lambda$. However, over a manifold with boundary,
the action is quasi-invariant: it transforms by a boundary term, $I_{\text{CS%
}} \to I_{\text{CS}} - (k/4\pi )\int_{\partial M} \text{Tr}(\lambda\, dA)$.
The Euler-Lagrange variation is

\begin{equation*}
\delta_{\text{EL}} I_{\text{CS}} = \frac{k}{4 \pi}\int_M 2 \text{Tr}(\delta
A \wedge F) + \frac{k}{4\pi} \int_{\partial M} \text{Tr}(\delta A \wedge
A)\, ,
\end{equation*}
where the bulk piece gives the zero curvature field equation.
Naively, the boundary piece can be dropped if $A =0$ or if $\delta A
= 0$, neither of which is a gauge invariant condition. There is no
local term at the boundary that can be added to the action to
restore the symmetry of the bulk theory. Boundary terms which are
not gauge invariant are unsatisfactory since, generically, they make
the Noether charges ill defined: under a gauge transformation that
is non trivial at infinity, the charges can take unbounded values
and may require ad-hoc regularizations as happens, for instance, in
CS gravity for asymptotically AdS spaces. This issue was addressed
in Refs. \cite{MOTZ1,MOTZ2}, where the CS action was supplemented by
introducing a second connection $\bar{A}$ and a boundary term that
turns
the action functional into a \textit{transgression form} \cite%
{Chern-74,Nakahara}. This expression is gauge invariant (and not quasi
invariant as is the case for the integral of a CS form in a manifold with
boundary). The transgression action reads

\begin{equation}  \label{transgression}
I[A,\bar{A}] = \int_{M} \mathcal{C}(A) - \int_{\bar{M}} \mathcal{C}(\bar{A})
- \int_{\partial M=\partial \bar{M}}\text{Tr} (A \wedge \bar{A}).
\end{equation}

Here the two connections $A$ and $\bar{A}$ have support on two different
manifolds with a common boundary where they interact. This functional is
invariant (modulo winding number) under independent gauge transformations
for $A$ and $\bar{A}$,

\begin{eqnarray}
A\rightarrow A^{\prime}&=& g^{-1}(A+d)g  \notag \\
\bar{A}\rightarrow \bar{A}^{\prime}&=& \bar{g}^{-1}(\bar{A}+d)\bar{g}\; ,
\end{eqnarray}
provided that the gauge transformations for both connections are the same at
the boundary, $g|_{\partial M}=\bar{g}|_{\partial M}$. The suitable
generalization of this idea when more than two manifolds meet at a common
boundary will be shown to be the appropriate scheme to describe multiple
bubbles. We shall pick up on this point again in section \ref{Boundary_sec}.

iv) For the gauge group $SO(2,2,\mathbb{R})$, there is the well
known interpretation for this action as equivalent to General
Relativity in $2+1$ dimensions with negative cosmological constant
\cite{Achucarro,Witten88} (at least perturbatively
\cite{Witten:2007kt}). Thus, the action for each wall can describe
the geometry of a pseudo-Riemannian surface of constant negative
curvature. This is the negative curvature space-time analogue of a
conventional double bubble, whose walls are surfaces of constant
positive curvature (according to the conjecture. This scheme offers
the possibility for matching different three-dimensional spacetime
geometries with a common boundary. This construction will be
discussed as an explicit example in section 3. As we will see, the
form of the interaction between the three geometries at the
intersection depends on how one chooses the surface terms.

Geometrically, some key features of the CS form can be understood by
considering the characteristic form quadratic in the field strength $P_4:=%
\text{Tr} (F\wedge F)$ (Chern character). This four form is closed and
therefore in a contractible open patch is exact, $P_4 = d\, \mathcal{C}$.
Chern-Simons theory on a closed manifold can in this way be interpreted as
the integral of a characteristic form over a manifold of one dimension
higher. For a concrete example, consider CS theory on a manifold which is
topologically a three-sphere $S^3$. This can be regarded as the boundary of
some four-manifold $B$. Let $A$ be a gauge field defined on $S^3$. Under
suitable topological assumptions, we can define an extension of this gauge
field to $B$, which for convenience we also call $A$. The CS action on $S^3$
is then equal to:

\begin{gather}  \label{Single_Bubble}
I_\text{CS}(S^3) \equiv \frac{k}{4\pi}\int_{B} \text{Tr}(F \wedge F)\, ,
\end{gather}

The extension of the fiber bundle $E(S^3)$ to a bundle $E(B)$ over the
four-dimensional space is a non-trivial matter\footnote{%
This extension to $E(B)$ can always be found when the cohomology group $%
H_3(BG,Z)$ of the classifying space $BG$ is trivial. Any bundle over a
manifold is the pullback bundle induced by the embedding of the manifold
into the classifying space. Therefore, if all three-cycles in the
classifying space are boundaries, there always exists a four manifold
bounded by $M_3$ such that an extension of the bundle onto the interior
exists. If $H_3(BG,Z)$ is non-trivial, the Chern-Simons theory may still be
defined\cite{Dijkgraaf-89} but the concept of an interior manifold may break
down.}. For example, gauge-related connections on $S^3$ may have different
extensions in the interior. So, the gauge field on $B$ cannot really be
regarded as completely fixed by the physical data on $S^3$. From the fact
that the Characteristic form defines an integer cohomology class, it follows
that the action will be gauge invariant, modulo some integer multiple of $%
2\pi$, provided the level $k$ is chosen to be an integer.

Since the characteristic form is closed, it defines a gauge theory intrinsic
to the three-sphere. Formally one could say that this describes the surface
dynamics of a bubble containing $B$ as its interior. This is not altogether
accurate because $B$ is not a fixed background. However it does suggest an
interesting idea, which we will now outline.


\section{Double bubble}


Let us consider a double bubble configuration like the one shown in Figure %
\ref{Double_Bubble_Diagram}. The double-bubble is made up of three different
manifolds, which we shall call the \emph{walls}, joined at a common
boundary. In the case of interest the walls are three-dimensional and they
meet on their two-dimensional intersection. This edge is a three-way
branching surface, so we have a structure which is not a Hausdorff manifold,
but rather a more general kind of cell complex. It is sometimes referred to
as a rectifiable set \cite{Almgren-66}, which means that it is arbitrarily
close in measure to being a manifold, with the singular set of non-manifold
points being of measure zero. The bubble complex can be regarded as the
union of boundaries of the interiors of the bubbles, i.e. the union of
boundaries of four-dimensional manifolds.

Let us now discuss two different approaches for constructing a Chern-Simons
foam or multiple bubble. The fist approach (section \ref{First_Method}) is
very natural from a four-dimensional point of view of bubble interiors, with
Characteristic forms living in them. This leads to two connections on each
wall. In the second approach (section \ref{Second_Method}), the walls
themselves play the prominent role (with the four-dimensional interiors
being reduced to a kind of metaphysical meaning). It is possible to define a
meaningful action for a single connection on each wall. In the rest of the
paper, we shall leave aside the first method and concentrate on the second.


\subsection{Action as sum of characters of 4D topological spaces}

\label{First_Method}

Inspired by equation (\ref{Single_Bubble}) for the single ``bubble" one can
postulate an action which is the sum of integrals of a Characteristic form
over each of the three four-dimensional interiors:

\begin{gather}  \label{Three_Characteristic_Forms}
I[A_1, A_2, A_3]=\frac{k_1}{4\pi}\int_{B_1} \text{Tr}(F_1 \wedge F_1) +
\frac{k_2}{4\pi} \int_{B_2} \text{Tr}(F_2 \wedge F_2) + \frac{k_3}{4\pi}
\int_{B_3} \text{Tr}(F_3 \wedge F_3)\, + [\text{Boundary terms}].
\end{gather}

\begin{figure}[t]
\par
\begin{center}
\includegraphics[height=3in]{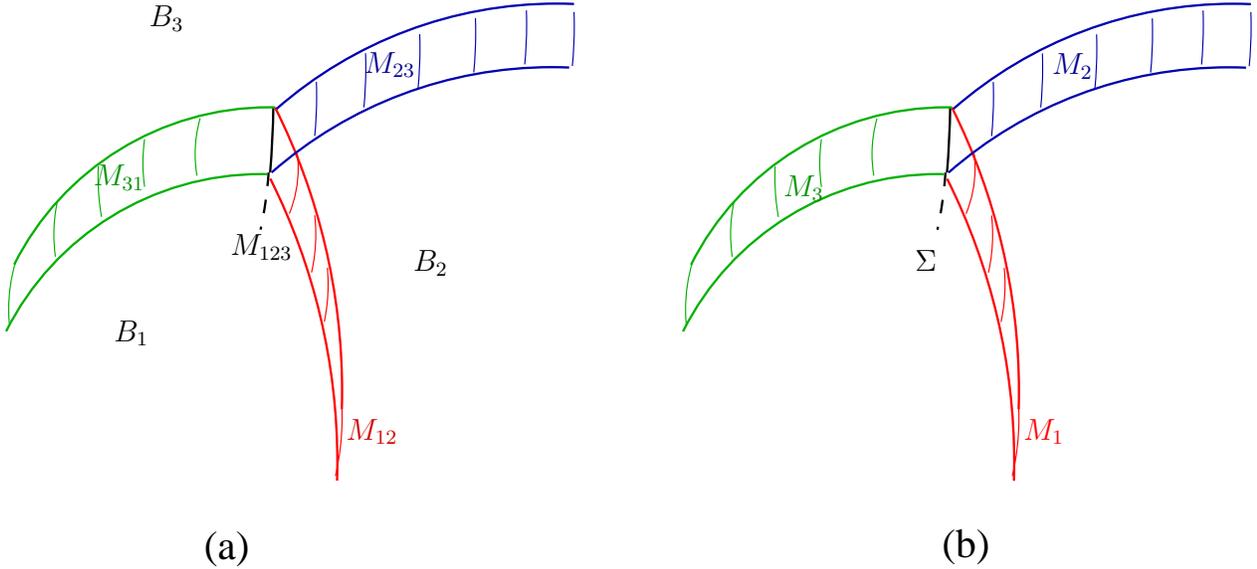}\vspace{-.15in}
\end{center}
\par
\caption{A slice of the double bubble shows three walls meeting at the
intersection. a) The first approach involves introducing a connection on
each four-dimensional bubble interior, $B_1$ and $B_2$ and the exterior
region $B_3$; b) The second approach, the main subject of this paper,
involves a single connection defined intrinsically on each of the walls $M_1$%
, $M_2$ and $M_3$.}
\label{Slice}
\end{figure}

There are three gauge fields, one in each of the four-dimensional regions $%
B_1$, $B_2$ and $B_3$, as shown in Figure \ref{Slice}.

One could neglect the exterior region, which would amount to fixing the
connection $A_3 \equiv 0$, but, it seems more appropriate to be democratic
and keep all three connections. Now, since each of the bulk terms in (\ref%
{Three_Characteristic_Forms}) is a closed form, it could be traded for a CS
form on the surface that encloses the respective four-volume, $B_i$. In this
way one defines an intrinsic CS theory on the walls and intersection of the
double bubble. The result would be the sum of three transgression forms
defined on the three bubble walls,

\begin{eqnarray}
I[A_1, A_2, A_3]&=&\frac{k_1}{4\pi}\int_{M_{12}} [ \mathcal{C}(A_1)-
\mathcal{C}(A_2)] + \frac{k_2}{4\pi} \int_{M_{23}} [ \mathcal{C}(A_2)-
\mathcal{C}(A_3)]  \notag \\
&& + \frac{k_3}{4\pi} \int_{M_{31}} [ \mathcal{C}(A_3)- \mathcal{C}(A_1)] +
\int_{M_{123}} [\text{Surface terms}].  \label{Three_CS_forms}
\end{eqnarray}

This functional depends on the difference between the CS forms obtained by
approaching each wall from both sides. The corresponding connections ($A_i$)
induced by their values on the neighboring volumes need not match. There
might be interesting cases in which this possibility can be useful. For
example, if the curvature two-form $F$ the same on each side of $M_{ij}$,
for in that case, the connections must differ at most by a gauge
transformation, $A_i= g^{-1}(A_j+d)g$ and the corresponding difference of CS
forms is a closed form describing a WZ theory at the two-dimensional
boundary.

The doubling of connections on each wall seems somewhat excessive and there
is no obvious interpretation of the fields. Even for three-dimensional AdS
gravity, it is not essential to introduce the difference of two CS forms:
there is an equivalent formulation with a single Chern-Simons form for the
AdS group (which can be generalised to higher odd dimensions). So we shall
not pursue this approach further in this article.

The action proposed above is one way of formally defining a foam as
embedded in an auxiliary four-dimensional manifold. The resulting
action is constructed with two CS forms in the three-dimensional
walls of the double bubble. This carries the disadvantage of having
two dynamically independent connection fields with same quantum
numbers defined on each three-surface. Alternatively, we may try a
different intrinsic definition, for a single field on each wall (and
intersections of them), without reference to any interior regions.


\subsection{Intrinsic theory on the bubble walls (Abelian Case)}

\label{Second_Method} 

The action proposed above is one way of formally defining a foam as
embedded in an auxiliary four-dimensional manifold. The resulting
action is constructed with two CS forms in the three-dimensional
walls of the double bubble. This carries the disadvantage of having
two dynamically independent connection fields with same quantum
numbers defined on each three-surface. Alternatively, we may try a
different intrinsic definition, for a single field on each wall (and
intersections of them), without reference to any interior regions.

To illustrate the construction, let us look at the simplest case of abelian
Chern-Simons theory. We will introduce an action which does not involve any
metric or conformal structure on the intersection and which preserves gauge
invariance (something which is not possible for a single manifold with
boundary, see section \ref{Boundary_sec}). We will see that this leads to a
consistent variational principle and therefore a sensible theory at least at
the classical level.


\subsubsection{The boundary coupling}


Consider three 3-manifolds $M_{i}$ all sharing the same boundary $\partial
M_{1} = \partial M_{2} = \partial M_{3} = \Sigma$. The edge $\Sigma$ is a
smooth two-dimensional space where the interaction takes place which is the
analogue of a vertex in a Feynman diagram for point particle interactions.
On each manifold $M_{i}$, a connection $A^{(i)}$ is defined. The action is
defined as a sum of the corresponding CS functionals with level $k_{i}$,

\begin{equation}  \label{3-transgression}
I[A^{(1)}, A^{(2)}, A^{(3)}] =\sum_{i=1}^3 \frac{k_{i}}{4\pi}\, \int_{M_{i}}
A^{(i)} \wedge dA^{(i)} + \int_{\Sigma} B[A^{(1)}, A^{(2)}, A^{(3)}] \, .
\end{equation}

This would be the correct generalization of the transgression (\ref%
{transgression}) if the boundary term were such that the functional be
invariant under independent gauge transformations on each $A^{(i)}$, subject
to the appropriate matching condition at the edge. Since the interaction
lagrangian must be a two-form, it can only be a sum of terms of the form $%
A^{(i)} \wedge A^{(j)}$.

In what follows we assume the levels $k_{i}$ to be all positive and allow
for an arbitrary sign in front of the kinetic term, $\epsilon_{i} = \pm 1$
to account for the sign of each level. The level can be eliminated from the
action by a suitable rescaling of the connections

\begin{gather*}
\hat{A}^{(i)} \equiv \sqrt{k_i} A^{(i)}\,
\end{gather*}

Thus, the most general possible action with interaction terms can be assumed
to be of the form,

\begin{gather}  \label{Tri_Intersection}
4 \pi I = \sum_{i=1}^{3}\epsilon_{i} \int_{M_{i}} \hat{A}^{(i)} \wedge d\hat{%
A}^{(i)} + \int_{\Sigma} [f_{1} \hat{A}^{(2)} \wedge \hat{A}^{(3)} + f_{2}
\hat{A}^{(3)} \wedge \hat{A}^{1} + f_{3} \hat{A}^{(1)} \wedge \hat{A}^{(2)}].
\end{gather}

The question is now, what restrictions are imposed on $\{f_{i},
\epsilon_{i}, k_{i} \}$ by the requirements of gauge invariance, and that $I$
should have a well posed variational problem.


\subsubsection{Gauge Invariance}

\label{Gauge_Section} 

Under independent gauge transformations of the different
connections, $\hat{A}^{(i)} \rightarrow
(\hat{A}^{\prime})^{(i)}=\hat{A}^{(i)} + d\hat{\lambda}^{(i)}$, the
action changes by

\begin{align*}
\delta I = - \int_{\Sigma} \Big[ & \left( \epsilon_{1} d\hat{\lambda}^{(1)}
+ f_{3} d\hat{\lambda}^{(2)} - f_{2} d\hat{\lambda}^{(3)}\right) \wedge \hat{%
A}^{(1)} \\
& + \left( \epsilon_{2}d\hat{\lambda}^{(2)} + f_{1} d\hat{\lambda}^{(3)} -
f_{3} d\hat{\lambda}^{(1)} \right) \wedge \hat{A}^{(2)} \\
& + \left( \epsilon_{3} d\hat{\lambda}^{(3)} + f_{2} d\hat{\lambda}^{(1)} -
f_{1} d\hat{\lambda}^{(2)} \right) \wedge \hat{A}^{(3)} \Big] \, .
\end{align*}

The right hand side of this equation vanishes identically for arbitrary $%
A^{(i)}$ provided the $\lambda^{(i)}$'s are such that

\begin{equation}  \label{matching}
\left[
\begin{matrix}
\epsilon_{1} & f_{3} & - f_{2} \\
- f_{3} & \epsilon_{2} & f_{1} \\
f_{2} & - f_{1} & \epsilon_{3}%
\end{matrix}
\right] \left[
\begin{matrix}
d\hat{\lambda}^{(1)} \\
d\hat{\lambda}^{(2)} \\
d\hat{\lambda}^{(3)}%
\end{matrix}
\right] =0 \,\,\, .
\end{equation}

The existence of non trivial solutions depends on making a reasonable choice
of coupling constants. In particular, demanding the vanishing of the
determinant requires

\begin{equation}  \label{ef=1}
\epsilon_{1}\epsilon_{2}\epsilon_{3} + \epsilon_{1} f_{1}^2 +
\epsilon_{2}f_{2}^2 + \epsilon_{3}f_{3}^2 =0 \, .
\end{equation}
On the other hand, (\ref{matching}) can be solved as the general vanishing
eigenvalue equation

\begin{gather}
(\bm{\eta} + \bm{f} ) \vec{\alpha} = 0\, ,  \label{BIA}
\end{gather}
where we have defined the matrices

\begin{gather*}
\bm{\eta} := \text{diag}(\epsilon_{1}, \epsilon_{2}, \epsilon_{3})\, ; \quad %
\bm{f} : = \left(%
\begin{matrix}
0 & f_{3} & -f_{2} \\
- f_{3} & 0 & f_{1} \\
f_{2} & -f_{1} & 0%
\end{matrix}%
\right)
\end{gather*}

First we note that since for any solution of (\ref{BIA}), $\vec{\alpha}^T(%
\bm{\eta} + \bm{f}) \vec{\alpha}=0$, and in view of the antisymmetry of $%
\bm{f}$, $\vec{\alpha}$ must satisfy $\epsilon_{1} (\alpha^{(1)})^2 +
\epsilon_{2} (\alpha^{(2)})^2 + \epsilon_{3} (\alpha^{(3)})^2 =0 $. This
could only occur for a nontrivial $\vec{ \alpha}$ if and only if $\bm{\eta}$
is an indefinite ``metric'', which without loss of generality we take as%
\footnote{%
One could have chosen either $\bm{\eta} = \text{diag} (+1,+1,-1)$ or $%
\bm{\eta} = \text{diag}(-1,-1,+1)$, but both cases are related by a global
reversal of sign convention for orientations.}

\begin{gather}
\bm{\eta} = \left(
\begin{matrix}
+1 & 0 & 0 \\
0 & +1 & 0 \\
0 & 0 & -1%
\end{matrix}
\right) \, .
\end{gather}

Equation (\ref{ef=1}) reduces to the requirement that the components
of a vector $\vec{f}$ lie on the surface of a hyperboloid of unit
space-like distance from the origin, $(f_{1})^2 + (f_{2})^2
-(f_{3})^2 = 1\, $. For later convenience, a general point on this
hyperboloid can be parametrised as

\begin{align}
f_{1} & = -\sin \Omega + \xi \cos \Omega  \notag \\
f_{2} & = \cos \Omega + \xi \sin \Omega  \label{fi-s} \\
f_{3} & = \xi  \notag
\end{align}

As already mentioned, the other consistency condition is that for every
nontrivial solution of (\ref{BIA}), $\vec{\alpha}$ must be null, that is $%
(\alpha^{(1)})^2 + (\alpha^{(2)})^2 - (\alpha^{(3)})^2 = 0$. Since the
vector $\vec{\alpha}$ is null, it is determined by (\ref{BIA}) up to an
arbitrary normalization constant and lies on the null cone through the
origin, $\vec{\alpha} = \alpha_0 \, \vec{n}$, where

\begin{equation*}
\vec{n}=\left(
\begin{matrix}
\cos \theta \\
\sin \theta \\
1%
\end{matrix}%
\right) \,.
\end{equation*}

Consistency and solvability of equation (\ref{BIA}) relate the angle $\theta$
iand the coupling constants $f_{(i)}$ and one finds

\begin{eqnarray}  \label{cos}
\theta = \Omega\, \, , \, \, \,\, \,\,\, \xi = \mbox{arbitrary }.
\end{eqnarray}

From the previous analysis, we conclude that the gauge transformations at
the intersection must be such that
\begin{align}
\hat{\lambda}^{(1)} =& \hat{\lambda}^{(3)} \cos \Omega\, ,
\label{strange_gauge1} \\
\hat{\lambda}^{(2)} =& \hat{\lambda}^{(3)} \sin \Omega \, ,
\label{strange_gauge2}
\end{align}
in order for the full action (\ref{Tri_Intersection}) be gauge invariant.
Reinstating the coupling constants, and imposing the consistency expressions
(\ref{fi-s}), the action becomes

\begin{gather*}
4 \pi I = \sum_{i=1}^{3} \epsilon_{i} k_i \int_{M_{i}} A^{(i)} \wedge d
A^{(i)} + \int_{\Sigma} \left[-\sqrt{k_2 k_3}\sin \Omega\ A^{(2)} \wedge
A^{(3)} + \sqrt{k_3 k_1}\ \cos\Omega\ A^{(3)} \wedge A^{(1)} \right] \\
+ \xi \int_\Sigma \left[ \sqrt{k_1 k_2}\ A^{(1)} \wedge A^{(2)} + \sqrt{k_2
k_3}\cos \Omega\ A^{(2)} \wedge A^{(3)} + \sqrt{k_3 k_1}\sin\Omega\ A^{(3)}
\wedge A^{(1)}\right] \, .
\end{gather*}

As we have seen, for the abelian CS theory, $k$ is somewhat of a phoney
coupling constant. It can always be set to $+ 1$ by using the rescaled
connection $\hat{A}$. But this leads to the rather \textit{ad hoc} matching
of gauge parameters, (\ref{strange_gauge1}) and (\ref{strange_gauge2}), at
the intersection. It is more natural to require that the ``true" gauge
parameters $\lambda^{(i)} = \hat{\lambda}^{(i)} / \sqrt{k_{i}}$ be
continuous at the intersection:

\begin{gather}  \label{lambdas}
\lambda^{(1)}= \lambda^{(2)} = \lambda^{(3)}
\end{gather}

This can be achieved by the nontrivial matching for the levels

\begin{gather}
\sqrt{k_{1}}=\sqrt{k_{3}}\cos \Omega \,,  \label{tension1} \\
\sqrt{k_{2}}=\sqrt{k_{3}}\sin \Omega \,,  \label{tension2}
\end{gather}%
\newline
which is reminiscent of the matching condition for the tension in a three
string junction \cite{Dasgupta:1997pu}. This relation can be written also
more suggestively as a \textquotedblleft conservation law\textquotedblright\
for the levels:

\begin{equation}  \label{konservation}
k_1+k_2=k_3 \;.
\end{equation}

Finally, the action reduces to

\begin{align}
4 \pi I & = k_1 \left( \int_{M_{1}} A^{(1)} \wedge d A^{(1)} - \int_{M_{3}}
A^{(3)} \wedge d A^{(3)} - \int_{\Sigma} A^{(1)} \wedge A^{(3)} \right)
\notag \\
& \quad + k_2 \left( \int_{M_{2}} A^{(2)} \wedge d A^{(2)} - \int_{M_{3}}
A^{(3)} \wedge d A^{(3)} - \int_{\Sigma} A^{(2)} \wedge A^{(3)} \right)
\notag \\
& \quad + \xi \sqrt{k_{1} k_{2} }\int_{\Sigma} \left[ A^{(1)} \wedge A^{(2)}
+ A^{(2)} \wedge A^{(3)} + A^{(3)} \wedge A^{(1)}\right] \, .
\label{Nice_action}
\end{align}

Note that the arbitrary coefficient $\xi $ multiplies a term that is gauge
invariant by itself (the origin of this term will be discussed in section %
\ref{Boundary_sec}). This term does not contribute to the field equations
either and can therefore be dropped from the classical action. The
transgression action (\ref{transgression}) is recovered for $k_{2}=0$ ($%
k_{1}=k_{3}$, and $\xi =0$), and therefore expression \ref{Nice_action} can
be regarded as a generalization of the concept of transgression for the case
of three manifolds with a common boundary. Note that, for $k_{2}\neq 0$, the
transgression is \emph{not} recovered by setting $A_{(2)}=0$. This is so
because the gauge invariance of the action depends upon the existence of all
three connections and setting one of them to zero is not a gauge invariant
statement.


\subsubsection{Matching conditions}


Now we consider the Euler-Lagrange equations for the action (\ref%
{Nice_action}). Extremising it with respect to independent variations of
each gauge field, under the gauge invariant matching conditions

\begin{gather}  \label{true_matching}
A^{(1)}|_\Sigma= A^{(2)}|_\Sigma = A^{(3)}|_\Sigma \, ,
\end{gather}
(where $|_\Sigma$ denotes the pullback on differential forms onto $\Sigma$)
on the edge, one obtains

\begin{gather}
F^{(i)} = 0
\end{gather}
\newline
on each wall. In other words, the action is stationary with respect to
arbitrary infinitesimal variations of the connections on each wall, provided
the connections are flat and match continuously at the edge. These matching
conditions are the same that guarantee an extremum for the transgression
action (\ref{transgression}).


\subsubsection{Comments}


$\bullet$ The matching condition of the gauge parameters (\ref{lambdas})
means that out of the possible U(1)$\times $U(1)$\times $U(1) gauge symmetry
on $\Sigma$ (the independent gauge transformations $A^{(i)} \to A^{(i)} +
d\lambda^{(i)}$), only a diagonal subgroup $U^d(1)$ is preserved. This is
the most that can be achieved without introducing extra fields. This is an
exact symmetry of the action and of course is a symmetry of the matching
conditions (\ref{true_matching}). \newline
$\bullet$ Consistency requires that the sign of one of the $\epsilon$'s must
be opposite to the other two (we have taken $\epsilon_3$ to be of opposite
sign to $\epsilon_1$ and $\epsilon_2$). We can think of these signs as
labeling ingoing and outgoing gauge fields. The consistency conditions
furthermore impose the conservation law (\ref{konservation}): the net
incoming level is always equal to the net outgoing level. \newline
$\bullet$ The matching conditions for $\lambda_{(i)}$ are insensitive to the
value of the ``coupling constant'' $\xi$. Indeed it is easy to check that
under (\ref{true_matching}), the last term in (\ref{Nice_action}) is gauge
invariant by itself. Likewise, the matching conditions for $k$ and $A$ are
insensitive to $\xi$. For this reason, at least in the classical context,
all choices of $\xi$ define the same physical theory. One can fix this
coefficient choosing, for instance, $\xi=0$. \newline
$\bullet$ The interaction term $A^{(1)} \wedge A^{(2)}$ between the two
ingoing connections is eliminated by choosing $\xi = 0$. \newline
$\bullet$ For the abelian theory, the choice $\lambda^{(1)} =\lambda^{(2)} =
\lambda^{(3)}$ is just a convenient option. It is rather a matter of choice
whether we put the nontrivial matching condition into the $k$'s or into the $%
\hat{\lambda}$'s and $\hat{A}$'s. In other words $\hat{A}$ is just as good a
connection as $A$. When we come to treat the non-abelian theory, however,
this will no longer be the case and ${A}$ should be regarded as the true
connection. Therefore, the nontrivial matching of the $k$'s is the preferred
interpretation.


\subsubsection{Example}


In order to understand the physical consequences of the relation among the
different gauge transformations, let us examine the case of a double bubble
of three-dimensional walls with three $U(1)$ connections. The field
equations imply that the connection on each 3-manifold is locally flat, $%
F^{(i)}=0$. For instance, a nontrivial locally flat connection could be
defined in a spacetime with a topological defect produced by a puncture on
the spatial section. Each bubble wall consists of a three-dimensional
spacetime manifold $M_{2+1} $ whose spacelike sections have the topology of
a disc with a removed point, $M_{2+1} = (D_2-\{0\}) \times \mathbb{R}$. The
action that describes a $U(1)$ connection in this 2+1 manifold is the sum of
two CS for 3 and 1 dimensions, respectively,

\begin{equation}
I[A]= k\int_{M^{2+1}}A \wedge dA +k^{\prime}\int_{M^{0+1}}A \, .
\end{equation}

This can also be written in a more familiar form as

\begin{equation}
I[A]= 2k\int_{M^{2+1}}\left[ \frac{1}{2}A \wedge dA - A\wedge j \right] \, ,
\end{equation}
where $j=q\, \delta^{(2)}(x,y) \, dx \wedge dy$ is the two-form current
density source produced by a (magnetic) point charge $q=-\frac{k^{\prime}}{2k%
}$. The field equation is $F=j$, and the classical solution takes the form

\begin{equation}
A = \frac{q}{2 \pi} d\phi \, .
\end{equation}

Now we want to put three connections of this sort defined on the three walls
of a double bubble. Requiring the action to be invariant under independent
gauge transformations of each connection --provided they respect (\ref%
{lambdas}) on the common boundary $\Sigma=S^1 \times \mathbb{R}$--, implies
the conservation law

\begin{equation}  \label{k's}
k_1+k_2=k_3 \;,
\end{equation}
and the matching condition for the $A$'s in this case becomes

\begin{equation}  \label{q's}
q^{(1)} = q^{(2)}= q^{(3)}.
\end{equation}

Since $q^{(i)}=-\frac{k^{\prime}_i}{2k_i}$, there is also a matching for the
$k^{\prime}$'s,

\begin{equation}
k^{\prime}_1+k^{\prime}_2=k^{\prime}_3 \;.
\end{equation}


\subsection{Non-abelian double bubble}


The previous analysis of the conditions at the intersection carries over
straightforwardly to the case of non-abelian Chern-Simons theory. There is
one subtlety associated with gauge invariance. In the abelian theory we
eliminated the coupling constants $k_{(i)}$ on the wall by rescaling the
gauge field. In the non-abelian theory the connection in each wall
transforms as

\begin{gather}  \label{Gauge}
A^{(i)} \to g_{(i)} ^{-1} A^{(i)}g_{(i)} + g_{(i)} ^{-1} d g_{(i)}\; ,
\end{gather}
\newline
which means that we cannot rescale the gauge field $\hat{A} \equiv \sqrt{ k}
A$ without modifying the gauge transformation correspondingly: $\hat{A}%
^{(i)} \to g_{(i)} ^{-1} \hat{A}^{(i)}g_{(i)} + \sqrt{k_i}g_{(i)} ^{-1} d
g_{(i)}$. So, it will be more convenient to keep the $k_i$'s explicitly in
the action.

It is natural to require that the field equations be invariant under
(\ref%
{Gauge}) where the gauge parameter $g(x)$ is globally defined over the
bubble complex, so that the gauge symmetry group is\footnote{%
One might instead try to demand instead invariance underindependent $g_i$'s.
This would give an enhanced symmetry $G \times G $ or $G \times G\times G$
at the intersection. It turns out that this is not possible without the
addition of extra fields. Indeed, if we wish to have all three gauge fields
truly interacting at the intersection, such an enhanced symmetry seems to be
unwanted.} $G$. We therefore require

\begin{gather}
g_{(1)}|_\Sigma = g_{(2)}|_\Sigma = g_{(3)}|_\Sigma \, .
\end{gather}
\newline
There is no other obvious way to relate the gauge parameters that has a
chance of being consistent with the field equations.

Considering infinitesimal gauge transformations leads to exactly the same
analysis as in section \ref{Gauge_Section}, and one is led to an action of
the form (\ref{Nice_action}),
\begin{align}
4 \pi I & = k_1 \left( \int_{M_1} \mathcal{C}(A^{(1)}) - \int_{M_3} \mathcal{%
C}(A^{(3)}) - \int_{\Sigma}\text{Tr}(A^{(1)} \wedge A^{(3)})\right )  \notag
\\
& \quad + k_2 \left( \int_{M_{2}} \mathcal{C}(A^{(2)}) - \int_{M_{3} }
\mathcal{C}(A^{(3)}) - \int_{\Sigma} \text{Tr}(A^{(2)} \wedge A^{(3)})
\right)  \label{Nice_action_Nonabelian} \\
& \quad + \xi \sqrt{k_{1} k_{2} }\int_{\Sigma} \text{Tr}\left( A^{(1)}
\wedge A^{(2)} + A^{(2)} \wedge A^{(3)} + A^{(3)} \wedge A^{(1)}\right) \, ,
\notag
\end{align}
where again, one can choose $\xi =0$. It can be easily checked that this is
invariant under finite gauge transformations, up to a winding number term,
as discussed in Appendix \ref{Quantisation_section}. Then, the Euler
variation of the action implies $F^{(i)} = 0$ on the walls, and the gauge
fields themselves obey the fairly unexciting relation at the intersection

\begin{gather}  \label{As_equal}
A^{(1)}|_\Sigma = A^{(2)}|_\Sigma = A^{(3)}|_\Sigma\, ,
\end{gather}
\newline
and again, the conservation rule $k_{1} + k_{2} = k_{3}$ applies. Thinking
of walls with negative $\epsilon$ as ``ingoing" and walls with positive $%
\epsilon$ as ``outgoing", we have the conservation law

\begin{gather}  \label{conservation}
(\sum k_i)_{\text{in}} = (\sum k_i)_\text{out}\, .
\end{gather}

This concludes the analysis for the three-way intersection. Next we
generalise to a four-way intersection which sets the general pattern for
intersections of higher order. We will show that the conservation law (\ref%
{conservation}) holds also in that case. Furthermore, we will find an
interesting formula which generalises (\ref{As_equal}).


\subsection{Example: matching (2+1)-dimensional black holes}


To illustrate the non-abelian three-way intersection, we consider
the anti-de Sitter gauge group $SO(2,2)$ (or $SO(3,1)$, for the
Euclideanised case), for which the Chern-Simons construction
describes (2+1)-dimensional gravity with negative cosmological
constant \cite{Achucarro,Witten88}. The anti-de Sitter connection is
$A=\omega ^{ab}J_{ab}/2+e^{a}J_{a3}/l$, $a=0,1,2$, and the bilinear
form is Tr$(J_{AB}J_{CD})\propto \epsilon _{ABCD}$. The contribution
to the action of each wall is therefore an Einstein-Hilbert term
with negative cosmological constant: $~(k_{i}/l_{i})\int_{\epsilon
_{i}M_{i}}(R^{ab}-(1/3l_{i}^{2})e^{a}\wedge e^{b})e^{c}\epsilon
_{abc}$.
Therefore, Newton's constant on each wall is\footnote{%
Here we use the convention by which the Einstein-Hilbert action is
$\frac{1}{%
2\pi G}\int \sqrt{-g}(R-2\Lambda )d^{3}x$.}%
\begin{equation*}
2\pi G_{i}=\frac{l_{i}}{k_{i}}\,.
\end{equation*}

The field equations on each wall is $F^{(i)}=0$, corresponding to
the
first-order formulation of Einstein's equations: $R^{ab}+\frac{1}{l^{2}}%
e^{a}\wedge e^{b}=0$, and $T^{a}=de^{a}+\omega _{\,\,b}^{a}\wedge
e^{b}=0$. We would like to analyse the possibility of matching four
static (2+1) black
holes. The metric on each wall has the form \cite{BTZ}%
\begin{equation*}
ds^{2}=f^{2}(r)dt_{E}^{2}+\frac{dr^{2}}{f^{2}(r)}+r^{2}d\phi ^{2}\,,
\end{equation*}%
where $f^{2}(r)\equiv r^{2}/l^{2}-\mu $ and $\mu =Gm$ where $m$ is
the mass of the black hole. The angle $\phi $ has period $2\pi $ and
the Euclidean time $t_{E}$ has period $\beta =2\pi l/\sqrt{\mu }$
which guarantees regularity of the Euclidean metric at $f=0$, with
$\beta ^{-1}$ being
interpreted as the black hole temperature\footnote{%
The $\mu =0$ solution is somewhat special. Making the coordinate
redefinition $z=1/r$ we get the Poincar\'{e} half space metric: $%
ds^{2}=z^{-2}(dt_{E}^{2}+dz^{2}+d\phi ^{2})$. $r=0$ is actually
located at conformal infinity and $t_{E}$ can be identified with any
period without producing a conical singularity. Therefore the zero
mass solution can have any temperature.}. On each wall, the vielbein
and spin connection read
\begin{gather}
\frac{e^{0}}{l}=\frac{f\,dt_{E}}{l},\qquad \frac{e^{1}}{l}=\frac{dr}{lf}%
,\qquad \frac{e^{2}}{l}=\frac{r\,d\phi }{l}\,, \\
\omega ^{01}=\frac{r\,dt_{E}}{l^{2}}\,,\qquad \omega ^{12}=-fd\phi
\,,\qquad \omega ^{20}=0\,.
\end{gather}

The matching of black-hole geometries takes place on a common intersection $%
\Sigma $ at constant radial coordinate, which can in principle take
a
different value, $r_{i}=a_{i}$, on each wall. The intersection surface $%
\Sigma $ has topology $S^{1}\times S^{1}$, and using invariance
under co-ordinate transformations the angular co-ordinates of the
four walls can be made to match, $\phi _{i}=\phi _{j}$ and
$t_{E}^{i}/\beta _{i}=t_{E}^{j}/\beta _{j}$. Because this space is
not a Hausdorff manifold, it is not clear how to relate the
co-ordinates $r_{i}$ differentiably, but
this is not a problem since the matching conditions are for the \emph{%
pullback} of $A$ onto $\Sigma $: the $A_{r}dr$ components (in this case $%
e^{1}/l$) do not contribute. This is crucial for the consistency of
the analysis.

Now, we consider the matching condition for the connection, which
for the triple intersection is simply $A^{(i)}|_{\Sigma
}=A^{(j)}|_{\Sigma }$. By matching $e^{0}/l$, $e^{2}/l$, $\omega
^{01}$ and $\omega ^{12}$ pulled back onto $\Sigma $ one gets:
\begin{equation}
a_{i}/l_{i}=a_{j}/l_{j}\,,\qquad \beta _{i}/l_{i}=\beta
_{j}/l_{j}\,,\qquad \mu _{i}=\mu _{j}\,.  \label{Matching
parameters}
\end{equation}%
It is natural to assume the proper length of the time cycle at the
boundary of each wall to be the same, that is, $\beta _{i}=$ $\beta _{j}$, which in turn requires $l$ to be a universal constant (%
$l_{i}=l_{j}$). In this way the induced metric on $\Sigma $ is
continuous and the temperatures all match. One is therefore matching
solutions which are in thermal equilibrium but have different Newton
constants. The quantization law $k_{i}=$ integer \cite{Zanelli-94}
holds for the triple intersection (see section
\ref{Quantisation_section}) and implies a corresponding quantization
of the Newton's constants. Since $m_{i}=\mu
k_{i}/l$, the masses are not all the same but, using $(\sum k)_{\text{in}%
}=(\sum k)_{\text{out}}$ , we find the conservation law
\begin{equation}
m_1 + m_2 = m_3\, .
\end{equation}
It is natural to assume that all the Newton constants are positive
and account for the signs $\epsilon _{i}$ by reversing the
orientation of $M_{3}$ with respect to $\Sigma $. If one takes this
view, one should match two interior regions $\rho _{1}\leq a_{1}$
and $\rho _{2}\leq a_{2}$ with one exterior $\rho _{3}\geq a_{3}$ or
vice versa. The above law then says that the sum of masses in the
interiors equals the mass in the exterior. The Newton constants
satisfy:
\begin{equation}\label{Newtons_double_trouble}
 \frac{1}{G_1} + \frac{1}{G_2} = \frac{1}{G_3}\, .
\end{equation}

The problem can be studied in a more generality as the matching of
hyperbolic manifolds along homeomorphic boundaries. Here we have
only matched Euclidean black holes without angular momentum. They
have the
topology of a solid torus and we have matched along a surface of topology $%
S^1 \times S^1$. It would be possible to also include Euclidean
black holes with angular momentum, since they have the same
topology\cite{Carlip-94}.


\section{Higher order intersections}


It is possible to construct an action for more than three films
meeting at a two-dimensional intersection. Unlike the three-way
intersection, it is possible to have more general matching
conditions than $A^{(i)}|_\Sigma = A^{(j)}|_\Sigma$, which leave the
connections less determined. It can be checked that as one goes to
higher order, the problem becomes less determined. From the point of
view of a static foam, this indefiniteness suggests some kind of
instability with respect to the more basic three-way intersection,
as occurs in soap bubbles. However, the higher order intersections
may be of interest in describing dynamical scattering of films. Here
we discuss explicitly only the case of four-way intersections.

\subsection{Four-way intersections}

Let us consider now the situation where four walls meet at a single
2-dimensional intersection surface $\Sigma$. As before, we choose the
orientations such that $\partial M_{i} = \Sigma$, assume all levels $k_i>0$,
and define $\epsilon_{i} = \pm 1$. A general ansatz for the interaction
term, without introducing other fields or preferred coordinates, will have 6
coupling constants $f_{ij}$

\begin{gather}
4 \pi I = \sum_{i = 1}^4 \epsilon_i k_{i}\int_{M_{i}} \mathcal{C}(A^{(i)}) +
\int_{\Sigma} \sum_{i<j}\ \sqrt{k_{i} k_{j}}\, f_{ij}\, \text{Tr}(A^{(i)}
\wedge A^{(j)})\, .
\end{gather}

Again, the condition of infinitesimal gauge invariance imposes an algebraic
constraint at the intersection that can be represented by the matrix equation

\begin{equation} \label{BI_4}
(\boldsymbol{\eta }+\boldsymbol{f})\vec{\beta}=0\,,
\end{equation}%
with $\boldsymbol{\eta }=\text{diag}(\epsilon _{1},\epsilon _{2},\epsilon
_{3},\epsilon _{4})$, $\boldsymbol{f}$ is the antisymmetric matrix with
entries $f_{ij}$, and $\beta ^{(i)}=\sqrt{k_{i}}\,D\lambda ^{(i)}$. By
considering $\vec{\beta}^{T}(\boldsymbol{\eta }+\boldsymbol{f})\vec{\beta}=0$
we find that $\sum_{i}\epsilon _{i}k_{i}(\beta ^{(i)})^{2}=0$ so that $%
\boldsymbol{\eta }$ cannot have Euclidean signature. The matching condition
for the connections will also be of the form (\ref{BI_4}), with $\beta
^{(i)}=\sqrt{k_{i}}\,A^{(i)}|_{\Sigma }$.


\subsection{Scattering of Chern-Simons films}

\label{Scattering_Section}

\begin{figure}[t]
\begin{center}
\includegraphics[height=2.2in]{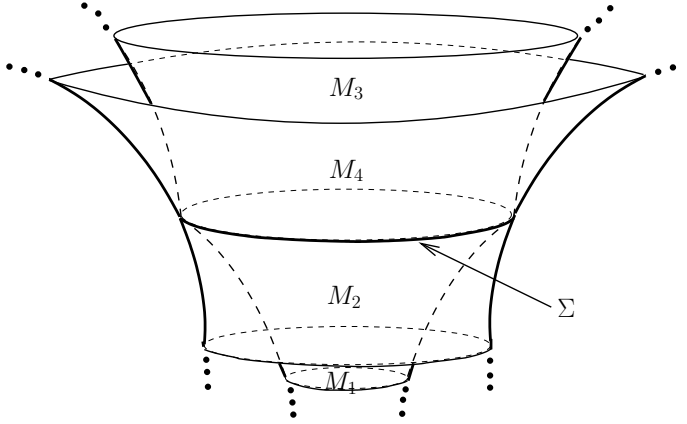}\\[0pt]
\end{center}\vspace{-.2in}
\caption{{\protect\small {The four way intersection may be
interpreted as a collision, with two incoming and two outgoing
walls.}}} \label{4BTZs}
\end{figure}

There is an interesting possibility which only occurs if
$\boldsymbol{\eta }$ has signature $(+,+,-,-)$, which corresponds to
the \textquotedblleft scattering" of two \textquotedblleft
incoming\textquotedblright\ and two \textquotedblleft
outgoing\textquotedblright\ walls (see fig. \ref{4BTZs}). In what
follows we restrict our attention to this case; situations with
three equal signs are similar in spirit to the three-way
intersection discussed above. The coupling constants $f_{ij}$ can be
chosen in such a way that the field equations give only two
independent equations for the $A^{(i)}$'s. We consider the $\bm{f}$
matrix of the form

\begin{equation*}
\boldsymbol{f}=\left(
\begin{array}{cc}
\emptyset  & -R^{T} \\
R & \emptyset
\end{array}%
\right)
\end{equation*}%
where $R$ and $\emptyset $ are $2\times 2$ matrices. Other forms of $%
\boldsymbol{f}$ can be obtained by an $SO(2,2)$ rotation from this case.
Then, the matching conditions

\begin{equation*}
\left(
\begin{array}{cc}
I & -R^{T} \\
R & -I%
\end{array}%
\right) \left(
\begin{array}{c}
\vec{\beta}_{\text{in}} \\
\vec{\beta}_{\text{out}}%
\end{array}%
\right) =0\,,
\end{equation*}%
reduce to two independent equations,

\begin{equation}\label{rotation}
\vec{\beta}_{\text{out}}\ =R\vec{\beta}_{\text{in}},
\end{equation}%
provided that $R^{T}R=I$ i.e. $R$ is an $O(2)$ matrix. We have defined $\vec{%
\beta}_{\text{in}}\ :=\left(
\begin{smallmatrix}
\sqrt{k_{1}}d\lambda ^{(1)}|_{\Sigma } \\
\sqrt{k_{2}}d\lambda ^{(2)}|_{\Sigma } \\
\end{smallmatrix}%
\right) $ and $\vec{\beta}_{\text{out}}\ :=\left(
\begin{smallmatrix}
\sqrt{k_{3}}d\lambda ^{(3)}|_{\Sigma } \\
\sqrt{k_{4}}d\lambda ^{(4)}|_{\Sigma } \\
\end{smallmatrix}%
\right) $. Now, by an appropriate choice of labels for the walls, $R$ may be
taken as an SO(2) matrix:
\begin{equation*}
R=\left(
\begin{array}{cc}
\cos \Omega & -\sin \Omega \\
\sin \Omega & \cos \Omega%
\end{array}%
\right) \,.
\end{equation*}%
Then, the action is
\begin{align}
4\pi I& =\sum_{i=1}^{4}\epsilon _{i}k_{i}\int_{M_{i}}\mathcal{C}(A^{(i)}) \\
& \quad -\int_{\Sigma }\left\{ \sqrt{k_{1}k_{3}}\,\text{Tr}(A^{(1)}\wedge
A^{(3)})+\sqrt{k_{2}k_{4}}\,\text{Tr}(A^{(2)}\wedge A^{(4)})\right\} \cos
\Omega  \notag \\
& \quad -\int_{\Sigma }\left\{ \sqrt{k_{1}k_{4}}\,\text{Tr}(A^{(1)}\wedge
A^{(4)})-\sqrt{k_{2}k_{3}}\,\text{Tr}(A^{(2)}\wedge A^{(3)})\right\} \sin
\Omega \,.  \notag
\end{align}

We assume the action to be invariant under gauge transformations that are
continuous at the intersection \footnote{%
For the abelian theory, this assumption is not necessary and so there is an
enhanced gauge symmetry $U(1)\times U(1)$ at the edge, the matching of the
gauge parameters being determined only by (\ref{rotation}).}, $%
g_{(1)}|_{\Sigma }=g_{(2)}|_{\Sigma }=g_{(3)}|_{\Sigma }=g_{(4)}|_{\Sigma }$%
. This implies
\begin{equation}\label{k_out}
\left(
\begin{array}{c}
\sqrt{k_{3}} \\
\sqrt{k_{4}}%
\end{array}%
\right) =\left(
\begin{array}{cc}
\cos \Omega & -\sin \Omega \\
\sin \Omega & \cos \Omega%
\end{array}%
\right) \left(
\begin{array}{c}
\sqrt{k_{1}} \\
\sqrt{k_{2}}%
\end{array}%
\right) \,.
\end{equation}%
As promised, the conservation law (\ref{conservation}) holds. One can use (%
\ref{k_out}) to express $\Omega $ as a function of the $k$'s
\begin{equation}
\Omega (k_{1},k_{2},k_{3},k_{4})=\tan ^{-1}\frac{\sqrt{k_{1}k_{4}}-\sqrt{%
k_{2}k_{3}}}{\sqrt{k_{1}k_{3}}+\sqrt{k_{2}k_{4}}}
\end{equation}%
and therefore $\Omega $ is not an independent coupling constant.

Eliminating $\Omega $ we can write
\begin{equation}\label{RMatrix}
R=\frac{1}{k_{1}+k_{2}}\left(
\begin{array}{cc}
\sqrt{k_{1}k_{3}}+\sqrt{k_{2}k_{4}} & -\sqrt{k_{1}k_{4}}+\sqrt{k_{2}k_{3}}
\\
\sqrt{k_{1}k_{4}}-\sqrt{k_{2}k_{3}} & \sqrt{k_{1}k_{3}}+\sqrt{k_{2}k_{4}}%
\end{array}%
\right)
\end{equation}%
with
\begin{equation}\label{4ks}
k_{1}+k_{2}=k_{3}+k_{4}\,.
\end{equation}%
The matching condition for the gauge field at $\Sigma $ is:
\begin{equation}\label{Akmatch}
\left(
\begin{array}{c}
\sqrt{k_{3}}\,A^{(3)}|_{\Sigma } \\
\sqrt{k_{4}}\,A^{(4)}|_{\Sigma }%
\end{array}%
\right) =R\ \left(
\begin{array}{c}
\sqrt{k_{1}}\,A^{(1)}|_{\Sigma } \\
\sqrt{k_{2}}\,A^{(2)}|_{\Sigma }%
\end{array}%
\right) \,.
\end{equation}

Finally, it may be helpful to re-express the matching conditions in the
form:
\begin{equation}\label{A3}
\left(
\begin{array}{c}
A^{(3)}|_{\Sigma } \\
A^{(4)}|_{\Sigma }%
\end{array}%
\right) =U\ \left(
\begin{array}{c}
A^{(1)}|_{\Sigma } \\
A^{(2)}|_{\Sigma }%
\end{array}%
\right) \,,
\end{equation}%
where the matrix
\begin{equation*}
U\equiv \frac{1}{k_{1}+k_{2}}\left(
\begin{array}{cc}
k_{1}+\sqrt{k_{1}k_{2}k_{4}/k_{3}} & k_{2}-\sqrt{k_{1}k_{2}k_{4}/k_{3}} \\
k_{1}-\sqrt{k_{1}k_{2}k_{3}/k_{4}} & k_{2}+\sqrt{k_{1}k_{2}k_{3}/k_{4}}%
\end{array}%
\right)
\end{equation*}%
satisfies the curious relation $\text{tr}U=1+\det U$. It can now be
seen that if the $k$'s are not all the same there are non-trivial
solutions with $A^{(i)}\neq A^{(j)}$.

This situation can be interpreted as describing the scattering of two
\textquotedblleft ingoing" and two \textquotedblleft outgoing" walls, $%
M_{1},\;M_{2}\rightarrow M_{3},\;M_{4}$. Given a set of levels ($%
k_{1},k_{2},k_{3},k_{4}$) satisfying (\ref{4ks}), one is free to
specify any
values of ($A^{(1)},A^{(2)}$) and equation (\ref{Akmatch}) (or equivalently (%
\ref{A3})) determines the outgoing values ($A^{(3)},A^{(4)}$). So we can
interpret our action as describing an elastic scattering process: what goes
out is completely determined by what comes in. The ingoing data is not
constrained, except by the bulk field equations $F^{(i)}=0$.


\section{Extensions}


\subsection{CS foam and Transgression forms}

\label{Boundary_sec} 

For a single Chern-Simons theory on a manifold with boundary, the presence
of the boundary usually breaks the symmetry drastically. One must impose
boundary conditions which break the gauge invariance, or else break the
diffeomorphism invariance on the boundary down to conformal invariance. A
standard construction would be to introduce a preferred complex structure
and add a boundary term $A_{z}A_{\bar{z}}$, leading to a two-dimensional
conformal field theory \cite{Moore}. Alternatively, one may try to modify
the action so as to preserve the topological nature of the theory. This was
considered in Refs. \cite{MOTZ1,MOTZ2}, by adding a second connection and a
boundary term so that the Lagrangian becomes a Transgression form:
\begin{equation*}
\mathcal{T}(A^{(1)},A^{(2)})\equiv \mathcal{C}(A^{(1)})-\mathcal{C}%
(A^{(2)})-d\text{Tr}(A^{(1)}\wedge A^{(2)})\,.
\end{equation*}%
This preserves gauge invariance, with $g_{(1)}|_{\partial
M}=g_{(2)}|_{\partial M}$, without introducing any fixed metric or conformal
structure on the boundary. Furthermore, since the gauge transformations only
need to be related \emph{at the boundary}, one can take the action
\begin{equation*}
\int_{M}[\mathcal{C}(A^{(1)})-\mathcal{C}(A^{(2)})]-\int_{\Sigma \equiv
\partial M}\text{Tr}(A^{(1)}\wedge A^{(2)})
\end{equation*}%
and \textquotedblleft pull apart" $M$ in the middle to produce a blister
shaped space $M_{1}\bigcup M_{2}\bigcup \Sigma $ where $\partial
M_{1}=\partial M_{2}=\Sigma $. The action
\begin{equation*}
S[A^{(1)},A^{(2)}]\equiv \int_{M_{1}}\mathcal{C}(A^{(1)})-\int_{M_{2}}%
\mathcal{C}(A^{(2)})-\int_{\Sigma }\text{Tr}(A^{(1)}\wedge A^{(2)})
\end{equation*}%
is still gauge invariant. Now $\Sigma $ is interpreted as the boundary of
our region of space and $A^{(2)}$ can be interpreted as a gauge field lying
beyond the boundary. In the case of AdS gravity, this transgression method
was shown to successfully regulate the charges, which would otherwise
require ad hoc counterterms.

It is natural to try to generalise this method to more than two connections.
An ansatz for the transgression action of a double-bubble would be a linear
combination:
\begin{equation*}
I=\alpha _{12}S[A^{(1)},A^{(2)}]+\alpha _{23}S[A^{(2)},A^{(3)}]+\alpha
_{31}S[A^{(3)},A^{(1)}]
\end{equation*}%
(It may be helpful think of this as the integral of
\begin{equation*}
\alpha _{12}\mathcal{T}(A^{(1)},A^{(2)})+\alpha _{23}\mathcal{T}%
(A^{(2)},A^{(3)})+\alpha _{31}\mathcal{T}(A^{(3)},A^{(1)})
\end{equation*}%
over the chain
\begin{equation*}
C=M_{1}+M_{2}+M_{3}
\end{equation*}%
provided we are careful that $A^{(i)}$ only has its support on the closure
of $M_{i}$.)

We note the identity
\begin{equation*}
\mathcal{T}(A^{(1)},A^{(2)})+\mathcal{T}(A^{(2)},A^{(3)})+\mathcal{T}%
(A^{(3)},A^{(1)})=-d\text{Tr}(A^{(1)}\wedge A^{(2)}+A^{(2)}\wedge
A^{(3)}+A^{(3)}\wedge A^{(1)})\,.
\end{equation*}%
which allows us to write
\begin{align*}
I& =(\alpha _{12}-\alpha _{31})S[A^{(1)},A^{(2)}]+(\alpha _{23}-\alpha
_{31})S[A^{(2)},A^{(3)}] \\
& \quad -\alpha _{31}\int_{\Sigma }\text{Tr}(A^{(1)}\wedge
A^{(2)}+A^{(2)}\wedge A^{(3)}+A^{(3)}\wedge A^{(1)})\,.
\end{align*}%
This is exactly the same as action (\ref{Nice_action}), which we had found
by the more constructive approach. What we have done can therefore be seen
as a generalisation of the transgression method. We start with a linear
combination of transgressions. Then the manifold with boundary is pulled
apart into \emph{three} cobordant manifolds.

The generalisation to more than three connections now seems clear:
Assign to a manifold with boundary a lagrangian $\sum_{i>j}\,\alpha
_{ij}\,\mathcal{T}(A^{(i)},A^{(j)})$; Then \textquotedblleft pull
apart" the manifold into a set of cobordant
manifolds, the walls $M_{i}$, so that each Chern-Simons form $\mathcal{C}%
(A^{(i)})$ is integrated over the corresponding wall $M_{i}$. This
is sufficient to guarantee gauge invariance. It is also sufficient
to guarantee that the consistent matching conditions:
$A^{(i)}|_{\Sigma }=A^{(j)}|_{\Sigma }\ \forall i,j$ will always
provide a solution. For certain choices of the coefficients $\alpha
_{ij}$ there may be more interesting matching conditions, as
happened in our example of four-way scattering discussed above.


\subsection{Quantisation of coupling constants}

\label{Quantisation_section}

So far we have neglected to mention the winding number contribution to the
gauge transformation. Under a general gauge transformation, the
double-bubble action (\ref{Nice_action_Nonabelian}) transforms by:
\begin{equation*}
-\frac{k_{1}}{24\pi }\int_{C_{13}}\text{Tr}\left( a^{-1}da\wedge \lbrack
a^{-1}da,a^{-1}da]\right) -\frac{k_{2}}{24\pi }\int_{C_{23}}\text{Tr}\left(
b^{-1}db\wedge \lbrack {b}^{-1}db,b^{-1}db]\right)
\end{equation*}%
where $C_{13}=M_{1}\cup \Sigma \cup (-M_{3})$ and $a$ is a continous
(strictly speaking it must be $C^{1-}$ so that $da$ has at most bounded
discontinuity at $\Sigma $) gauge parameter which coincides with $g_{1}$ in $%
M_{1}$ and $g_{3}$ in $M_{3}$, and similarly for $C_{23}$. Note that it is
possible, by choosing a non-Hausdorff topology on the double-bubble\footnote{%
This is the branching universe topology discussed in Ref. \cite{Visser}.},
to treat $C_{13}$ and $C_{23}$ as closed Hausdorff sub-manifolds, so that
these integrals make sense mathematically. Therefore, we can immediately
deduce that the double bubble action transforms under a gauge transformation
by
\begin{equation*}
2\pi k_{1}n_{1}+2\pi k_{2}n_{2}
\end{equation*}%
with $n_{1}$ and $n_{2}$ integer winding numbers. So the argument regarding
quantisation of coupling constants applies to the double bubble just the
same as to a closed manifold.

For more general types of bubble network, presumably the quantisation
argument should also apply but a less crude proof is required. (It may help
to think in terms of cohomology of Chern-Simons forms on chains rather than
manifolds \cite{Chern-74}.)

\subsection{Some mathematical subtleties}


In the context of topological field theory \cite{Atiyah-90} one encounters
various situations which are defined on graphs or other non-manifold
structures. So the present discussion is not new in that sense. In what
sense is our approach different?

A common approach is to assign some \textquotedblleft colour" or
combinatorial information to the different pieces of a graph. For example,
in the Ponzanno-Regge theory \cite{Ponzano-Regge}, an irreducible
representation of SU(2) is assigned to each of the edges on the skeleton of
some triangulated manifold. In another, more recent, proposal \cite%
{quantum_graphity}, the states are related to the adjacency matrix of
sub-graphs. In contrast to these approaches, we introduced an action
depending on local fields on the walls. In this, it is similar to Plateu's
problem, in which one is interested in defining the volume form over bubble
networks. Minimising this volume gives a discrete set of solutions, with
simple algebraic rules like (\ref{Double_Bubble_Formula}) for the
intersections.

It is true that our model also gives discrete degrees of freedom on the
walls and algebraic relations like (\ref{konservation}), (\ref{q's}) and (%
\ref{Newtons_double_trouble}) for the intersections, but this is an
accident of CS theory in three dimensions. We have seen that an
action involving local fields and their derivatives can be well
defined. So in principle one can generalise to a theory which has
local degrees of freedom on the walls. For many theories one may run
into trouble because the space is not a Hausdorff
manifold\footnote{%
A space with branching is either: i) a Hausdorff topological space which is
not a manifold or; ii) a manifold which is not Hausdorff, depending on how
one defines the topology \cite{McCabe-05}.} and therefore the derivative of
the fields across the intersections is not defined. In a second order field
theory, like the Klein-Gordon system, it seems that the situation is
hopeless because one encounters the second normal derivative of the field;
likewise for 3 dimensional gravity in the second order metric formalism. In
the case of CS theory we encounter only a first normal derivative which, due
to Stokes' theorem, does not cause problems. There should be no problems in
generalising to higher-dimensional CS theories, which do have local degrees
of freedom \cite{BGH}, and perhaps to other theories such as GR in its
first-order formalism and generalisations thereof \cite{Zumino,Mardones}.


\subsection{Bubbles in a different number of dimensions}


The procedure for defining an action for Chern-Simons bubbles is not special
to three dimensions. CS theory is defined in all odd dimensions, where one
could also expect to find other interesting theories of bubbles. It can be
seen that, assuming the form of the action to be given like in (\ref%
{Nice_action_Nonabelian}) as a sum of transgressions, the relation (\ref%
{sum-ks}) holds for all higher dimensions as well.

The simplest example of a foam would be a one-dimensional theory for the
abelian group $U(1)$. A bubble complex made of one-dimensional CS pieces can
be defined as follows. Let $M_{i}$ be a collection of 1-dimensional open
manifolds. We introduce the co-ordinate $s_{i}$ such that the manifold is
given by the line interval $s_{i}\in (0,1)$. The boundary of $M_{i}$ is thus
$\{s_{i}=1\}-\{s_{i}=0\}$. To each manifold we assign a single U(1)
connection $A^{(i)}$. The bubble complex is then made by joining pieces
together at the boundaries in an arbitrary way. Writing the connection $1$%
-form as $A=a(s)ds$, the one-dimensional Chern-Simons action for a triple
intersection reads
\begin{equation*}
I[A^{(1)},A^{(2)},A^{(3)}]=\sum_{i=1}^{3}\epsilon
_{i}k_{i}\int_{M_{i}}A^{(i)}\,.
\end{equation*}%
In this case, there is no need of a boundary term at the vertex. The
condition of gauge invariance under independent gauge transformations of
each $A^{(i)}$ that is continuous at the vertex, implies
\begin{equation}
\sum_{i=1}^{3}\epsilon _{i}k_{i}=0.  \label{sum-ks}
\end{equation}%
Again this implies that $\boldsymbol{\eta }$ is an indefinite
\textquotedblleft metric\textquotedblright .


\section{Summary}


Chern-Simons theory is normally defined on a manifold, but here we have
argued that the theory is also perfectly well defined on a cell complex,
such as a network of bubbles, which is not a Hausdorff manifold.

The powerful requirement of gauge invariance determines the action and gives
the conservation law for the levels at each intersection:
\begin{equation*}
(\sum k)_{\text{out}}=(\sum k)_{\text{in}}
\end{equation*}%
where \textquotedblleft in" or \textquotedblleft out" refer to those walls
for which the Chern-Simons term comes with a plus or a minus sign,
respectively (with respect to the orientation $\partial M_{i}=+\Sigma $).

The action principle, with unconstrained independent variation of
all gauge fields at the intersection, leads to matching conditions
for the connection. In a CS theory for a $U(1)$ connection in the
presence of an electromagnetic point source, this matching
establishes a relation among the charges. For an $SO(2,2)$
connection in $2+1$ dimensions, the black hole masses are related.

We found evidence of a qualitative distinction between branching
intersections, with only one \textquotedblleft incoming" wall, and
scattering intersections, with multiple ingoing and outgoing walls.
In the case of a three-way branching the condition was that the
connections match. For the example of three walls containing
charges, the condition was that all three charges are equal. In the
case of a four-way scattering, a more general matching condition
(\ref{Akmatch}) is obtained.

Several connections to similar an possible related systems seem to deserve
further study. In particular, the curious similarity between the matching
for the levels (\ref{tension1},\ref{tension2}) and the tension in
intersecting $D$-branes as in \cite{Dasgupta:1997pu} or \cite%
{Townsend:1996em}, in which the Chern-Simons couplings, through requirements
of charge conservation, seem to play an important role in determining what
kinds of intersections are allowed. There is also some similarity between
our work and the Chern-Simons membranes considered in refs \cite{Mora_thesis}
and \cite{Edelstein-08}, and with work on discontinuous connections and
Euler densities \cite{Gravanis-03} .
\newline

 \textbf{Acknowledgements:} We thank A. Giacomini and E.
Gravanis for many useful comments. S.W also thanks Nikolaos
Mavromatos for a very useful discussion. The Centro de Estudios
Cient\'{\i}ficos (CECS) is funded by the Chilean Government through
the Millennium Science Initiative and the Centers of Excellence Base
Financing Program of Conicyt. CECS is also supported by a group of
private companies which at present includes Antofagasta Minerals,
Arauco, Empresas CMPC, Indura, Naviera Ultragas and Telef\'{o}nica
del Sur. S.W. would like to thank staff at CPT, University of Durham
for kind hospitality during the final stages of this work. This work
was supported in part by Fondecyt grants 1061291 and 1085323.


\end{document}